\documentstyle[prl,aps,preprint]{revtex}
\newcommand{\tr}{{\rm tr}}
\newcommand{\Tr}{{\rm Tr}}
\newcommand{\Dirac}{{\rm D}}

\newcommand{\thru}[1]{\mathrel{\mathop{#1\!\!\!/}}}
\newcommand{\eq}[1]{eq.~(\ref{eq:#1})}

\newcommand{\D}{D}
\newcommand{\Da}{\hat{D}}

\newcommand{\bfx}{\mbox{\boldmath $x$}}

\newcommand{\scom}{\mbox{\textsf{\textit{c}}}}
\newcommand{\sj}{\mbox{\textsf{\textit{j}}}}
\newcommand{\sm}{\mbox{\textsf{\textit{m}}}}

\newcommand{\sv}{\mbox{\textsf{\textit{v}}}}

\newcommand{\sC}{\mbox{\textsf{\textit{C}}}}
\newcommand{\sD}{\mbox{\textsf{\textit{D}}}}
\newcommand{\sDa}{\hat{\sD}}
\newcommand{\sF}{\mbox{\textsf{\textit{F}}}}

\newcommand{\sX}{\mbox{\textsf{\textit{X}}}}

\newcommand{\mlr}{m_{LR}}
\newcommand{\mrl}{m_{RL}}

\begin{document}

\draft
\tighten
\def\footnoterule{\kern-3pt \hrule width\hsize \kern3pt}

\title{Derivative expansion for the effective action of
chiral gauge fermions. The normal parity component.}

\author{L.L. Salcedo}

\address{
{~} \\
Departamento de F\'{\i}sica Moderna \\
Universidad de Granada \\
E-18071 Granada, Spain
}

\date{\today}
\maketitle

\thispagestyle{empty}

\begin{abstract}
Explicit exact formulas are presented, up to fourth order in a strict
chiral covariant derivative expansion, for the normal parity component
of the Euclidean effective action of even-dimensional Dirac
fermions. The bosonic background fields considered are scalar,
pseudo-scalar, vector and axial vector. No assumptions are made on the
internal symmetry group and, in particular, the scalar and
pseudo-scalar fields need not be on the chiral circle.
\end{abstract}

\vspace*{1cm}

\vspace*{0.2cm} Keywords:\ \ chiral fermions, fermion determinant,
derivative expansion, effective action, gauge field theory, anomaly

\vspace*{\fill}

\section{Introduction}
\label{sec:1}

The effective action plays an important role in quantum field theory,
both from the phenomenological and from the formal points of view,
since it compactly embodies the renormalized properties of the system;
Green's functions, $S$-matrix elements and the expectation values of
observables can be extracted from it. In addition it is suitable to
study non-perturbative issues. Consequently the effective action
functional has been the subject of a very large amount of work from
different points of view, which include its proper definition and
renormalization, its symmetries and their anomalous breaking, its
properties in curved space-times or finite temperature, and its
calculation. An excellent and quite complete review for chiral
fermions in flat space-time and zero temperature, the case of interest
to us, can be found in \cite{Ball:1989xg}. For anomalies in curved
space-times see e.g. \cite{Alvarez-Gaume:1985dr}. Work on chiral
fermions at finite temperature can be found in
\cite{Pisarski:1996ne,Salcedo:1998tg}.

In the present work and its companion paper \cite{Salcedo:2000II} we
concentrate on the computational issues. Concretely we deal with the
calculation of the effective action of Dirac fermions in the presence
of bosonic external fields, thus the functional adds one-loop Feynman
diagrams with fermions running on the loop and bosonic external
legs. We restrict ourselves to the case of even dimensions (since
there are some technical differences with the odd-dimensional case),
zero temperature and flat space-time. The class of external bosonic
fields to be included is that of scalar, pseudo-scalar, vector and
axial vector fields. Coupling to higher tensor fields is not included,
nevertheless, the class considered here is quite large since no
assumption will be made on the internal symmetry group, that is, the
external fields are arbitrary matrices free from algebraic assumptions
regarding their dimension, commutativity, chiral circle constraint or
any other constraints among them.

The computation of the effective action functional in closed form is
not possible in general, and thus several asymptotic expansions have
been devised. In the heat kernel expansion the terms are classified by
its scale dimension, that is, each term has a well defined number of
external fields and derivative operators. It is the computationally
simplest expansion and so it has been carried out to considerably
large orders even for curved space-times \cite{vandeVen:1998pf}. Other
expansions can be regarded as resummations of this one. In the
perturbative expansion, the contributions are classified by the number
of external legs, that is, the number of external fields in the term,
and all orders in their momenta are added. This is a weak field
approximation which however captures the non-locality of the exact
functional. On the other hand, in the covariant derivative expansion,
to be considered in this work, the contributions are classified by the
number of chiral covariant derivatives, or equivalently, by the number
of Lorentz indices. Both definitions are equivalent in the absence of
external tensor fields. This counting is appropriate for external
fields with a smooth space-time dependence and weak gauge fields. The
scalar and pseudo-scalar fields need not be weak, and this allows to
study non-perturbative issues such as spontaneous symmetry
breaking. Two important properties of this expansion is that chiral
symmetry is preserved (modulo anomalies) separately for each term, and
that the terms are local.

The effective action is a ultraviolet divergent quantity which needs
to be renormalized. The non-perturbative definition of this functional
in the general chiral case is not so straightforward as for
vector-like theories (in which there is no pseudo-scalar nor axial
fields) due to the presence of essential chiral anomalies which affect
the imaginary part (the phase of the fermionic determinant)
\cite{Leutwyler:1985em}. Nevertheless such a definition exists
\cite{Ball:1989xg} and so the effective action is perfectly
well-defined also in the chiral case, displaying only the standard
renormalization ambiguities in the form of polynomial terms. The
subtleties in the non-perturbative definition of the effective action
are much alleviated within the asymptotic expansions noted above. In
particular, within the covariant derivative expansion to be worked out
in this paper for the real part of the effective action and in
\cite{Salcedo:2000II} for the imaginary part, there is only a finite
number of ultraviolet divergent terms. They are afflicted by
polynomial ambiguities (including anomalies and multivaluation) but
are otherwise unique. All higher order terms are ultraviolet finite
and so unambiguous.

It should be emphasized that the covariant derivative expansion of the
effective action functional, although asymptotic, is perfectly well
defined, that is, it does not depend on how it is written or
computed. (This is not true for other expansions such as the
commutator expansion, as we will discuss below.) This is because it
corresponds to classify the terms by their scaling under (covariant)
dilatations of the external fields. Each of these terms is a well
defined and universal functional, in the sense that they hold for any
possible internal symmetry group, and the same is true for the
perturbative or heat kernel expansions. For these two latter
expansions it is relatively easy to write an explicit form for each
term without putting restrictions on the internal symmetry group. The
reason for this is that all non commutative quantities, namely the
matrix-valued external fields, are treated perturbatively and each
term contains only a finite number of them. On the other hand, in a
strict covariant derivative expansion the gauge fields are treated
perturbatively but the scalar and pseudo-scalar fields appear to all
orders in every single term, and these two fields do not commute with
each other. Nevertheless, in this work and in \cite{Salcedo:2000II} we
show that the terms of the derivative expansions are also fully
amenable to explicit computation in closed form without assuming
particular properties of the internal symmetry group. In addition, no
chiral rotation (or diagonalization) is required to express these
universal functionals. Our result takes an analytical form in terms of
the external fields.

In this work we deal with the real part of the effective action and
compute it until fourth order in the covariant derivative expansion
for arbitrary even space-time dimensions. The real part is the
simplest one since it is free from essential chiral anomalies and
multivaluation. It contains only a scale anomaly. In
\cite{Salcedo:2000II} the study of the imaginary part is dealt with at
leading order in the derivative expansion for two- and
four-dimensional space-times. Besides the results themselves, in this
paper and in \cite{Salcedo:2000II} we introduce notational conventions
which are very well suited to the chiral problem. In particular, we
find that the formulas for the real part in the full chiral case are
identical to those of the vector-like case.

In Section \ref{sec:2} we introduce our notational conventions, some
of which are not standard, and also introduce the effective action. In
Section \ref{sec:5} we show how, for the real part of the effective
action, an appropriate notation allows to reduce the full chiral case
to the vector-like case, and also to carry out explicitly all
integrations over the momentum in the fermionic loop. The Section ends
with explicit formulas for the real part of the effective action up to
four covariant derivatives, based in the convenient method introduce
by Chan for bosons \cite{Chan:1986jq}. Finally in Section \ref{sec:3}
we illustrate the meaning of the formulas by analyzing the case of
second order and two space-time dimensions and a particular case is
worked out explicitly. Next we show how our notation allows to obtain
commutator expansions quite efficiently, and the Section is ended by
giving the analogous explicit formulas for bosons, once again without
restriction on the internal symmetry group.

\section{General considerations}
\label{sec:2}

\subsection{The Dirac operator}
\label{subsec:II.A}

The Euclidean effective action of fermions in a $d$-dimensional flat
space ($d$ even) is $\int d^dx\bar\psi\Dirac\psi$, where $\Dirac$ is
the Dirac operator. The class of operators to be considered is, in
terms of the left-right (LR) fields,
\begin{equation}
\Dirac= \thru{\D}_R P_R +\thru{\D}_LP_L + \mlr P_R + \mrl P_L \,,
\end{equation}
where $P_{R,L}=\frac{1}{2}(1\pm\gamma_5)$ are the projectors
on the subspaces $\gamma_5=\pm 1$. Our conventions are
\begin{equation}
\gamma_\mu=\gamma_\mu^\dagger\,,\quad \{\gamma_\mu,\gamma_\nu\}=
2\delta_{\mu\nu}\,,\quad
\gamma_5=\gamma_5^\dagger=\gamma_5^{-1}=
\eta_d\gamma_0\cdots\gamma_{d-1}\,,\quad
\tr_{\rm Dirac}(1)= 2^{d/2}\,.
\end{equation}
With $\eta_d=\pm i^{d/2}$ (a concrete choice will not be needed).
$\D^{R,L}_\mu= \partial_\mu+v^{R,L}_\mu$ are the chiral covariant
derivatives. The external bosonic fields $v^{R,L}_\mu(x)$ and $\mlr
(x)$, $\mrl (x)$ are matrices in internal space (referred to as
flavor), the identity in Dirac space and multiplicative operators in
$x$ space. Unitarity of the theory imposes restrictions on the
hermiticity properties of these fields, namely, $v_{R,L}$ must be
antihermitian and $\mlr^\dagger=\mrl$. In practice these restrictions
will play almost no role in the calculation\footnote{However we will
exploit the fact that the spectrum of $m^2_R(x)$ (defined below in
\eq{n6}) is non negative to choose the branch cuts of the logarithm or
squared root functions along the negative real axis.} and $\mlr$ and
$\mrl$ will be treated as independent variables. To avoid infrared
divergences we will assume that $\mlr$ and $\mrl$ are nowhere singular
matrices.

In terms of the vector-axial (VA) variables
\begin{equation}
\Dirac= \thru{\D}_V+\thru{A}\gamma_5+ S+{\gamma_5}P\,,
\end{equation}
where $\D^V_\mu=\partial_\mu+V_\mu$ is the vector covariant derivative
and
\begin{equation}
v_{R,L}= V\pm A\,,\quad \mlr = S+P\,,\quad \mrl = S-P\,.
\end{equation}

\subsection{Chiral transformations}

Chiral transformations act as follows
\begin{eqnarray}
&& \D^{R}_\mu \to \Omega_{R}^{-1}\D^{R}_\mu\Omega_{R}\,, \quad  
\D^{L}_\mu \to \Omega_{L}^{-1}\D^{L}_\mu\Omega_{L}\,,  
\nonumber \\
&& \mlr \to \Omega_L^{-1} \mlr \Omega_R\,,
\quad  \mrl \to \Omega_R^{-1} \mrl \Omega_L\,,
\label{eq:34}
\end{eqnarray}
where $\Omega_{R,L}(x)$ are independent, nowhere singular and
otherwise arbitrary matrices in flavor space.  Vector gauge
transformations correspond to chiral transformations in the diagonal
subgroup $\Omega_R=\Omega_L$, i.e. $\Dirac\to
\Omega^{-1}\Dirac\Omega$.

It will be convenient to introduce the two combinations
\begin{equation}
m^2_R= \mrl\mlr\,,\quad  m^2_L= \mlr\mrl \,,
\label{eq:n6}
\end{equation}
which transform solely under $\Omega_R$ or $\Omega_L$
respectively. Note that the two matrices $m^2_{R,L}(x)$ are related by
a similarity transformations and thus they have the same spectrum.

The different pieces in the Dirac operator transform in a well defined
manner under chiral transformations, namely, the quantities $\mlr $,
$\mrl $, $\D_R$ and $\D_L$ fall in the chiral representations $LR$,
$RL$, $RR$ and $LL$ respectively, cf. \eq{34}. New objects with well defined
chirality are obtained by multiplication in the natural way, i.e., if
$X_{ab}$ falls in the representation $ab$ and $Y_{bc}$ in $bc$, for
$a,b,c=R,L$, the product $Z_{ac}=X_{ab}Y_{bc}$ (no sum over $b$ is
implied) falls in the representation $ac$. If in addition $X_{ab}$ is
a multiplicative operator, its chiral covariant derivative
\begin{equation}
(\Da_\mu X)_{ab}= \D^a_\mu X_{ab}- X_{ab}\D^b_\mu\,, \quad a,b = R,L
\end{equation}
is also multiplicative. In particular, 
\begin{eqnarray}
\Da_\mu\mlr:=(\Da_\mu m)_{LR}=\partial_\mu \mlr + v_\mu^L \mlr  -\mlr v_\mu^R
\,,\nonumber\\
\Da_\mu\mrl:=(\Da_\mu m)_{RL}=\partial_\mu \mrl + v_\mu^R \mrl  -\mrl v_\mu^L 
\,.
\label{eq:c8}
\end{eqnarray}
In addition, 
\begin{equation}
F^R_{\mu\nu}= [\D^R_\mu,\D^R_\nu]\,,\quad F^L_{\mu\nu}= [\D^L_\mu,\D^L_\nu]\,,
\label{eq:c9}
\end{equation}
are also chiral covariant and multiplicative. All multiplicative
chiral covariant local objects come as combinations of $m$, $F$ and
their chiral covariant derivatives.

\subsection{The effective action}
The fermionic effective action $W=-\log\int {\cal D}\bar\psi{\cal
D}\psi\exp(-\int d^dx\bar\psi\Dirac\psi)$ is given by
\begin{equation}
W[v,m]=-\Tr\,\log(\Dirac)\,.
\end{equation}
As is well-known this expression is formal due to the presence of
ultraviolet divergences. Mathematically proper definitions of
$\Tr\,\log(\Dirac)$ exist in the literature. Here it is only necessary
to emphasize that different definitions of $W$ may at most differ by
terms which are local polynomial of dimension $d$, that is,
polynomials in the external fields $v$ and $m$ and their
derivatives. This is a standard result of perturbative quantum field
theory that can be established by isolating the ultraviolet divergent
one-loop Feynman graphs, which may contain at most $d$ insertions, and
expanding on the external momenta to extract the divergent part. In
practice this means that any method consistent with the formal
expression can be used to make $W$ finite, since it will give the same
ultraviolet finite contributions as any other method. The effective
action of a concrete physical system described by the Dirac operator
$\Dirac$ will be given by any of the renormalized versions of $W$ plus
an appropriate local polynomial counterterm.

As usual, it will be convenient to introduce the pseudo-parity
transformation, $R\leftrightarrow L$ (that is, $v_L\leftrightarrow
v_R$ and $\mlr\leftrightarrow\mrl$ or equivalently $A_\mu\to -A_\mu$,
$P\to -P$) and split the effective action into its even and odd
components under this transformation, $W^\pm$, i.e.,
\begin{equation}
W[v,m] = W^+[v,m] + W^-[v,m] \,.
\end{equation}
(Also known as normal and abnormal parity components.)  Due to parity
invariance, which involves and additional $(x_0,\bfx)\to(x_0,-\bfx)$
in the fields, the pseudo-parity odd component of the effective action
$W^-[v,m]$ is that containing the Levi-Civita pseudo-tensor. In
addition, $W^\pm[v,m]$ coincide with the real and imaginary parts,
respectively, of the (Euclidean) effective action when the standard
hermiticity for the fields is assumed.

By computing the effective action within a derivative expansion, we
mean expressing it as a sum of terms with a well-defined number of
covariant derivative operators but any number of scalars. This is
equivalent to classify the terms by the number of Lorentz indices they
contain. We emphasize this since sometimes the expression ``derivative
expansion'' is used in the literature to denote large mass expansions
or expansions in the number of all kind of commutators, that is, those
implied by the covariant derivative plus those of the form $[m^2,\ ]$,
etc. Further, we mean writing each of the terms using building blocks
which are multiplicative operators in $x$ space (i.e., functions of
$x$ and not differential operators) and trivial in Dirac space (since
the Dirac trace can be explicitly computed for any given order in a
derivative expansion). The trace in flavor space will not be worked
out since we are not imposing any algebraic constraint on the flavor
structure of the fields.

Note that, because $d$ is even and all Lorentz invariants are formed with
$\delta_{\mu\nu}$ and $\epsilon_{01\dots d-1}$, there are no odd-order
terms in the derivative expansion.

In this work we will consider the pseudo-parity even component of the
effective action which is simpler due to its lack of chiral
anomalies. The pseudo-parity odd component is worked out in
\cite{Salcedo:2000II}.

\subsection{Local basis in flavor space}
\label{subsec:II.D}
Occasionally it will be useful to diagonalize the flavor matrices
$\mlr(x)$ and $\mrl(x)$. This can be done by
solving the eigenvalue problem (at each point $x$)
\begin{equation}
\left( \matrix{ 0 & \mrl \cr \mlr & 0 } \right)
\left( \matrix{ \hfill |j,R\rangle \cr \pm|j,L\rangle } \right)
=
\pm m_j\left( \matrix{ \hfill |j,R\rangle \cr \pm|j,L\rangle } \right)
\,
\end{equation}
which yields
\begin{eqnarray}
\mlr|j,R\rangle &=& m_j|j,L\rangle\,,\quad \mrl|j,L\rangle =m_j|j,R\rangle\,,
\nonumber \\
\langle j,L|\mlr &=& m_j\langle j,R|\,,\quad \langle j,R|\mrl = 
m_j\langle j,L|\,.
\label{eq:35}
\end{eqnarray}
$\langle j,R|$ is the dual basis of $|j,R\rangle$, $\langle
j,R|k,R\rangle=\delta_{jk}$ (no orthonormality of the basis is
implied). The numbers $m_j^2$ are the common eigenvalues of $m^2_R$
and $m^2_L$ and $|j,R\rangle$ and $|j,L\rangle$ are their
eigenvectors. Because $\mlr^\dagger=\mrl$, the eigenvalues $m_j$ can
be taken to be positive.

\subsection{Some notational conventions}

In this section we will introduce some notational conventions which
are essential for carrying out the subsequent calculations. They are
also used in \cite{Salcedo:2000II}.

Because $W^+$ is by definition invariant under the exchange of the
labels $R$ and $L$, each term $T$ in the expansion of $W^+$ will have
a pseudo-parity conjugate term, $T^*$, obtained from $T$ by exchanging
everywhere the labels $R$ and $L$. (Note that due to the cyclic
property, it may actually happen that $T$ and $T^*$ coincide.) Thus
we will use the following convention:
\begin{description}
\item[Convention 1.] In $W^+$, the terms $T$ and $T^*$ will be
identified, so that under this Convention $T$ actually stands for
$\frac{1}{2}(T+T^*)$.
\end{description}
It is important to note that word ``term'' is used here in a very
specific way, namely, it refers only to contributions of the form
$T=\tr(X)$, that added produce the effective action. (By extension,
term may denote also the quantity $X$ itself. In what follows two such
quantities $X$ and $Y$ differing only by a cyclic permutation will be
identified, since they are equivalent inside the trace.) Of course,
the identification implied in Convention 1 does not apply to smaller
pieces or factors inside each term.

Due to chiral covariance each term is a product of factors with
well-defined chirality (namely, $LR$, $RL$, $RR$ or $LL$) correctly
combined to preserve chirality (i.e., $\cdots X_{ab}Y_{bc}\cdots$). In
addition, due to the cyclic property of the trace, if a term starts
with label $a=R,L$ it has to end also with same label $a$,
e.g. $\tr(X_{RR}Y_{RL}Z_{LR})$.  Based on these observations we will
make the following convention:
\begin{description}
\item[Convention 2.] In expressions where the chiral labels are
combined preserving chirality, these labels are redundant and will be
suppressed, thus a term such as $\tr(X_{RR}Y_{RL}Z_{LR})$ will be
written as $\tr(\textsf{\textit{XYZ}})$. Note that
$\tr(\textsf{\textit{XYZ}})$ could be expanded either as
$\tr(X_{RR}Y_{RL}Z_{LR})$ or $\tr(X_{LL}Y_{LR}Z_{RL})$, but both
expressions are equivalent under the Convention 1. We will choose the
first label as $R$.
\end{description}
For instance
\begin{eqnarray}
\tr(\sF_{\mu\nu}\,\sDa_\mu\sm\,\sDa_\nu\sm) &=&
\tr(F_{\mu\nu}^R \Da_\mu m_{RL} \Da_\nu m_{LR})
\nonumber \\
 &=&
\frac{1}{2}\tr(F_{\mu\nu}^R \Da_\mu m_{RL} \Da_\nu m_{LR}) +
\frac{1}{2}\tr(F_{\mu\nu}^L \Da_\mu m_{LR} \Da_\nu m_{RL})\,.
\end{eqnarray}

This notation allows to use the objects $\sm$, $\sv_\mu$, etc, as
ordinary operators (i.e., elements of an associative algebra).  For
instance, the property $\sm f(\sm)=f(\sm)\sm$ is verified as is
readily checked. With this notation eqs. (\ref{eq:c8}) and
(\ref{eq:c9}) become
\begin{equation}
\sDa_\mu\sm= \sD_\mu\sm- \sm\sD_\mu = [\sD_\mu,\sm]\,,\quad
\sF_{\mu\nu}= [\sD_\mu,\sD_\nu]\,,
\end{equation}
and \eq{35} becomes 
\begin{equation}
\sm|\sj\rangle = m_j|\sj\rangle\,,\quad
\langle \sj|\sm = m_j\langle \sj|\,.
\end{equation}

Another essential property, the cyclic property of the trace, holds
too. For instance,
\begin{equation}
\tr(\sX\sm)= \tr(X_{RL}m_{LR})= \tr(m_{LR}X_{RL})=
\tr(m_{RL}X_{LR})= \tr(\sm\sX)\,.
\label{eq:10}
\end{equation}
Let us note that these conventions have to be slightly modified to
include the pseudo-parity odd component $W^-[m,v]$. In particular,
the cyclic property of the trace is modified and this fact is at the
origin of the chiral anomaly in this formalism\cite{Salcedo:2000II}.

Below, it will be necessary to carry out parametric integrations where
the parameters appear in different places tied to operators which do
not commute. To do these integrations another convention is needed
regarding the order of operators. Consider for instance the integral
\begin{equation}
\int\frac{d^dp}{(2\pi)^d}
\tr\left[ \frac{\sm}{p^2+\sm^2}\sv_\mu \frac{\sm}{p^2+\sm^2}\sv_\mu\right]\,.
\label{eq:n17}
\end{equation}
It can be computed by first taking the trace using the basis of
eigenvectors of $\sm$ introduced in subsection \ref{subsec:II.D}. This
yields an expression involving matrix elements of $\sv_\mu$ in these
basis, namely,
\begin{equation}
\int\frac{d^dp}{(2\pi)^d}\,
\sum_{j,k}
 \frac{m_j}{p^2+m_j^2}(\sv_\mu)_{jk}  \frac{m_k}{p^2+m_k^2}
(\sv_\mu)_{kj}
\,,
\end{equation}
where $(\sv_\mu)_{jk}=\langle j,L| v_\mu^L|k,L\rangle$ and
$(\sv_\mu)_{kj}=\langle k,R| v_\mu^R|j,R\rangle$. (For definiteness we
have assumed that the operator inside the trace in \eq{n17} has been
expanded as a $RR$ term.) The point of taking matrix elements is that,
since now everything is in terms of commuting numbers, the momentum
integral can be done explicitly. Instead of that, we will use an
equivalent but preferable procedure, namely, we will label the
matrices $\sm$ according to their position relative to
$\sv_\mu$. Since there are two $\sv_\mu$ there are three possible
relative positions which are labeled by 1, 2 and 3. With this
prescription, the same integral can be represented unambiguously by
the formula
\begin{equation}
\int\frac{d^dp}{(2\pi)^d}\,
\tr\left[
 \frac{\sm_1}{p^2+\sm_1^2}\frac{\sm_2}{p^2+\sm_2^2}
\sv_\mu^2
 \right]\,.
\end{equation}
The usefulness of this notation is that, since the ordering is given
by the labels, $\sm_1$ and $\sm_2$ are effectively c-numbers and the
momentum integration can be carried out straightforwardly.
\begin{description}
\item[Convention 3.] In an expression $f(A_1,B_2,\dots)XY\cdots$ the
ordering labels $1,2,\dots$ will denote the actual position of the
operators $A,B,\dots$ relative the fixed elements $X,Y,\dots$ so that
$A$ is to be placed before $X$, $B$ between $X$ and $Y$, etc. That
is, for a separable function $f(a,b,\dots)=\alpha(a)\beta(b)\cdots$,
the expression stands for $\alpha(A)X\beta(B)Y\cdots$
\end{description}
Note that the fixed elements $X,Y,\dots$ appear only as simple factors
(i.e. perturbatively) whereas the labeled operators $A,B,\dots$ may
appear with a non perturbative functional dependence.  Also, note that
several different labeled operators can have the same label provided
that they commute.\footnote{The separation into fixed elements and
labeled operators is a matter of convenience. It is also possible to
label all operators to indicate their relative position and this is
often useful in order to carry out algebraic manipulations with a
computer.} Finally, in an expression with $n$ fixed elements inside
the trace, the labels $1$ and $n+1$ are equivalent, due to the cyclic
property of the trace. So, in the previous example, $\sm_3=\sm_1$. We
remark that Convention 3 is independent of Conventions 1 and 2.

The usefulness of this notation can be further exposed through the
following observation regarding commutators. Let $X$ be a single fixed
element and let the symbol $D_A$ denote the operation $[A,\ ]$, thus
it immediately follows from Convention 3 that $D_A= A_1-A_2$, that is
\begin{equation}
f(D_A)X= f(A_1-A_2)X\,.
\label{eq:n25}
\end{equation}
For instance,
\begin{equation}
e^A X e^{-A}  = e^{A_1}e^{-A_2}X= e^{A_1-A_2}X= e^{D_A}X\,,
\end{equation}
which is a well-known identity. (Alternatively, this example can be
regarded as a proof of \eq{n25}.) This observation will prove useful
to carry out commutator expansions (cf. subsection \ref{subsec:4.a}).

As another illustration of Convention 3, consider the following
identity, where $f(A)$ depends on the operator $A$ and $\delta A$
represent some first order variation of it,
\begin{equation}
\delta f(A)= \frac{f(A_1)-f(A_2)}{A_1-A_2}\delta A \,.
\label{eq:dfA}
\end{equation}
Here $\delta A$ is the fixed element referred to in the Convention and
$A_1$ and $A_2$ refer to $A$ before and after $\delta A$,
respectively. This identity can be proven as
follows.\footnote{Alternatively it can be proven by starting from
$\delta (A^{-1})= -A^{-1}\delta A A^{-1} = -A_1^{-1}A_2^{-1}\delta A$
and applying it to $f(A)=(2\pi i)^{-1}\int_\Gamma dz f(z)/(z-A)$ where
$\Gamma$ is positively oriented and encloses the spectrum of $A$, or
also it can be established by considering functions of the form
$f(x)=x^n$.} As is well-known
\begin{equation}
\delta e^{s A}=  \int_0^s dt\, e^{t A}\delta A e^{(s-t)A} \,,
\end{equation}
where $s$ is a c-number parameter. Using Convention 3, this can be
rewritten as
\begin{equation}
\delta e^{s A}=  \int_0^s dt e^{t A_1+ (s-t)
A_2}\delta A = \frac{e^{s A_1}-e^{s A_2}}{A_1-A_2}\delta A \,,
\end{equation}
where in the last step we have used that $A_1$ and $A_2$ are commuting
quantities. This identity is then generalized for arbitrary $f(A)$ by
Fourier transform. From another point of view, noting that $A_1-A_2$
is equivalent to $[A,\ ]$ for any $A$, \eq{dfA} is equivalent to
$[A,\delta f(A)]= [f(A),\delta A]$, which is a trivial consequence of
$\delta[A,f(A)]=0$.

Inside a trace the two ordering labels $1$ and $2$ become identical
due to the cyclic property, therefore \eq{dfA} yields
\begin{equation}
\Tr\left(\delta f(A)\right)= 
\Tr\left(f^\prime(A)\delta A \right) \,.
\end{equation}
($f^\prime$ denotes the derivative of $f$.)  The corresponding formula
for two successive first order variations, is
\begin{equation}
\Tr\left(\delta^\prime\delta f(A)\right)= 
\Tr\left(f^\prime(A)\delta^\prime\delta A 
+ 
\frac{f^\prime(A_1)-f^\prime(A_2)}{A_1-A_2} \delta^\prime A \delta A
\right) \,,
\end{equation}
which in particular implies
$\Tr\left([\delta^\prime,\delta]f(A)\right)=
\Tr\left(f^\prime(A)[\delta^\prime,\delta]A\right)$.

\section{Calculation of $W^+$ to fourth order}
\label{sec:5}

\subsection{Reduction to a vector-like theory}

A first benefit of Conventions 1 and 2 is that the functional
$W^+[v,m]$ will be formally identical to the effective action of a
vector-like theory. That is, if $W_{\text{V}}[V,S]$ denotes the
effective action functional when $A_\mu=P=0$,
\begin{equation}
W^+[v,m] = W_{\text{V}}[\sv,\sm]\,,
\end{equation}
provided that $W^+[v,m]$ and $W_{\text{V}}[V,S]$ have been
renormalized preserving chiral and vector gauge invariances
respectively. This is because when the functional $W^+$, expressed in
our notation in terms of $\sv$ and $\sm$, is particularized to the
vector-like case (i.e., $v_R=v_L$, $\mlr=\mrl$) no simplification
occurs; it remains unchanged in our notation. ($W^-$ vanishes for
vector-like configurations.) As a consequence, $W^+$ for the general
case can be obtained by computing only the effective action for
vector-like configurations. 

It is an essential point of this discussion that we are considering
field configurations which are free of constraints in flavor space:
the calculation of $W^+[v,m]$ through $W_{\text{V}}[V,S]$ requires
$V_\mu$ and $S$ to be generic in flavor space for the formal
replacements $V\to\sv$ and $S\to\sm$ in the formulas to be
well-defined. For instance, if the calculation of $W_{\text{V}}[V,S]$
were carried out in the particular case of commuting $V_\mu$ and $S$,
$[V_\mu,S]$ would be identified with zero and this would result in an
ambiguity in $W^+[v,m]$ by terms $[\sv_\mu,\sm]$, however
$[\sv_\mu,\sm]_{RL}=[V_\mu,\mrl]+\{A_\mu,\mrl\}$, which does not
vanish even for Abelian flavor groups, due to the axial
term. Therefore the proper procedure is first to compute
$W_{\text{V}}[V,S]$ in general, then make the replacements $V\to\sv$
and $S\to\sm$, and afterwards particularize the resulting formulas to
the case at hand. This replacement is a kind of analytical
continuation from the vector-like to the chiral case.

\subsection{Chan's Method}

Here we present the most efficient method to compute $W^+$. Further
heuristic considerations are made in Section \ref{sec:3}. We have
already reduced the problem to a vector-like one, with effective
operator $\Dirac=\thru{\sD}+\sm$. Next, the theory is reduced to a
bosonic one using $\Dirac^\dagger\Dirac$ as operator. This procedure
is standard.
\begin{equation}
\Dirac=\thru{\sD}+\sm\,,\quad \Dirac^\dagger=\gamma_5
\Dirac\gamma_5 = -\thru{\sD}+\sm \,.
\end{equation}
The second equality implies that $\Dirac$ and $\Dirac^\dagger$ are
related by a similarity transformation and so they can be assigned the
same determinant, thus
\begin{equation}
W^+[v,m]=-\frac{1}{2}\Tr\log\left(\Dirac^\dagger \Dirac\right)\,.
\label{eq:38}
\end{equation}
Two remarks are in order. First, the notation $\Dirac^\dagger$ assumes
the fields have the standard hermiticity, but only the fact that
$\Dirac^\dagger$ is similar to $\Dirac$ is essential. Second, within a
given particular renormalization prescription (e.g. $\zeta$-function)
\eq{38} will be correct modulo local polynomial counterterms.  This is
entirely sufficient since, as noted above, the action only determines
the effective action modulo a local polynomial of dimension $d$.

The bosonic operator $\Dirac^\dagger \Dirac$ can be worked out to give
\begin{eqnarray}
\Dirac^\dagger \Dirac &=&
-\sD_\mu^2-\frac{1}{2}\sigma_{\mu\nu}\sF_{\mu\nu}-[\thru{\sD},\sm]+\sm^2\,,
\nonumber \\
&=& P^2+U\,,
\end{eqnarray}
where $\gamma_\mu\gamma_\nu=\delta_{\mu\nu}+\sigma_{\mu\nu}$, and
\begin{equation}
P_\mu= \sD_\mu\,,\quad P^2=-P_\mu^2\,, \quad 
U= \sm^2-\gamma_\mu\sDa_\mu\sm-\frac{1}{2}\sigma_{\mu\nu}\sF_{\mu\nu}\,.
\label{eq:37}
\end{equation}
(Note that $P^2$ is unrelated to the pseudo-scalar field
$P=(\mlr-\mrl)/2$.)

A manifestly gauge invariant form for the derivative expansion of the
effective action of a general bosonic theory with Klein-Gordon
operator of the form $P^2+U$ has been given in \cite{Chan:1986jq} to
fourth order and in \cite{Caro:1993fs} to sixth order.  Those formulas
assume only that $P_\mu$ is a covariant derivative, i.e., of the form
$\partial_\mu+W_\mu(x)$, and $U(x)$ a Lorentz-scalar field. The fields
$W_\mu(x)$ and $U(x)$ are matrices in some internal space. To fourth
order the result is\footnote{A misprint in the formula presented in
\cite{Chan:1986jq} was corrected in \cite{Caro:1993fs}.}
\begin{eqnarray}
\Tr\log(P^2+U) &=&
\int\frac{d^dx\,d^dp}{(2\pi)^d}\tr\Big[
-\log(N)
+\frac{p^2}{d}N_\mu^2
\nonumber \\ &&
-\frac{2p^4}{d(d+2)}\left(
-2N_\mu^4+(N_\mu N_\nu)^2+2(NN_{\mu\mu})^2
+ 4NF_{\mu\nu}NN_\mu N_\nu 
+(F_{\mu\nu}N^2)^2 
\right)
\nonumber \\ &&
+\cdots
\Big]\,,
\label{eq:36}
\end{eqnarray}
where the following quantities have been defined
\begin{equation}
N=\frac{1}{p^2+U}\,,\quad N_\mu=[P_\mu,N]\,,\quad 
N_{\mu\mu}=[P_\mu,[P_\mu,N]]\,,
\quad
F_{\mu\nu}=[P_\mu,P_\nu]\,.
\end{equation}
In addition, $N_\mu^4$ stands for $(N_\mu^2)^2$, and $(N_\mu
N_\nu)^2=N_\mu N_\nu N_\mu N_\nu$, etc. The symbol $\tr$ includes all
internal degrees of freedom. The dots refer to terms with six or more
$P_\mu$. All terms are multiplicative operators (i.e. just functions
of $x$) and manifestly gauge invariant. Because all terms are
multiplicative, the trace cyclic property applies without
restrictions. In what follows, terms related by a cyclic
permutation will be identified when they appear inside the trace.

Eq. (\ref{eq:36}) can straightforwardly be applied to compute $W^+$
with the identifications
\begin{equation}
N_\mu=\sDa_\mu N\,,\quad N_{\mu\mu}=\sDa_\mu^2 N\,,\quad F_{\mu\nu}=
\sF_{\mu\nu}\,,
\end{equation}
and $N$ and $U$ given in \eq{37}. It is just necessary to note that in
the bosonic case $U$ is assumed to be of zeroth order, however, in the
fermionic case $U$ contains terms of first and second order in
$\sD_\mu$, so $N$ has to be reexpanded in order to have a well-defined
derivative expansion of $W^+$:
\begin{eqnarray}
N^{-1}&=& \Delta+U^{(1)}+U^{(2)}\,,
\nonumber \\
\Delta &=& p^2+\sm^2\,,\quad
U^{(1)}=-\gamma_\mu \sDa_\mu \sm\,, \quad
U^{(2)}= -\frac{1}{2}\sigma_{\mu\nu}\sF_{\mu\nu}\,.
\end{eqnarray}

Before proceeding, two comments can be made. First,
$\Dirac\Dirac^\dagger$ can be used instead of $\Dirac^\dagger
\Dirac$. It corresponds to the replacement $\sm\to -\sm$ in the
previous expressions, so $W^+$ is an even functional of $\sm$. This is
readily verified in the final expression. Second, $\Dirac^2$ would
also produce another acceptable bosonic operator
${P^\prime}^2+U^\prime$, with
$P^\prime_\mu=\sD_\mu+\gamma_\mu\sm$. However, this form is not
suitable for a derivative expansion because $\gamma_\mu\sm$ is of
zeroth order. This implies that low orders in the derivative expansion
of the fermionic problem would pick up contributions from all orders
in the bosonic derivative expansion.

Combining the previous equations, and after taking the Dirac trace,
the following result is obtained:\footnote{We will use the notations
$W^+_n[v,m]$ and $W^+_{n,d}[v,m]$ where $n$ specifies the order in the
derivative expansion and $d$ the space-time dimension.}
\begin{eqnarray}
W^+_0[v,m] &=& -\frac{2^{d/2}}{2}\int\frac{d^dx\,d^dp}{(2\pi)^d}\tr
\log\Delta
\label{eq:43} \\
W^+_2[v,m] &=&
-\frac{2^{d/2}}{2}\int\frac{d^dx\,d^dp}{(2\pi)^d}\tr\left[
-\frac{1}{2}\left(\frac{1}{\Delta}\sm_\mu\right)^2
+\frac{p^2}{d}\left(\frac{1}{\Delta^2}(\sm^2)_\mu\right)^2
\right]
\label{eq:44} \\
W^+_4[v,m] &=&
-\frac{2^{d/2}}{2}\int\frac{d^dx\,d^dp}{(2\pi)^d}\tr\Bigg[
-\frac{1}{2}\left(\frac{1}{\Delta}\sm_\mu\right)^4
+\frac{1}{4}\left(\frac{1}{\Delta}\sm_\mu\frac{1}{\Delta}\sm_\nu\right)^2
+\frac{1}{\Delta}\sm_\mu\frac{1}{\Delta}\sm_\nu\frac{1}{\Delta}\sF_{\mu\nu}
+\frac{1}{4}\left(\frac{1}{\Delta}\sF_{\mu\nu}\right)^2
\nonumber \\ &&
+\frac{p^2}{d}\Bigg\{
2\left(\frac{1}{\Delta^2}(\sm^2)_\mu\frac{1}{\Delta}\sm_\nu\right)^2
+2\frac{1}{\Delta^2}(\sm^2)_\mu\frac{1}{\Delta^2}\sm_\nu
\frac{1}{\Delta}(\sm^2)_\mu\frac{1}{\Delta}\sm_\nu
\nonumber \\ &&
+2\frac{1}{\Delta}(\sm^2)_\mu\frac{1}{\Delta^2}(\sm^2)_\mu
\frac{1}{\Delta}\sm_\nu\frac{1}{\Delta^2}\sm_\nu
+4\left(\frac{1}{\Delta^2}(\sm^2)_\mu\right)^2
\left(\frac{1}{\Delta}\sm_\nu\right)^2
\nonumber \\ &&
-4\frac{1}{\Delta^2}(\sm^2)_\mu\frac{1}{\Delta}\sm_\nu
\frac{1}{\Delta^2}\sm_{\mu\nu}
-4\frac{1}{\Delta^2}(\sm^2)_\mu\frac{1}{\Delta^2}\sm_\nu
\frac{1}{\Delta}\sm_{\mu\nu}
+\left(\frac{1}{\Delta^2}\sm_{\mu\nu}\right)^2
\Bigg\}
\nonumber \\ &&
-\frac{2p^4}{d(d+2)}\Bigg\{
-2\left(\frac{1}{\Delta^2}(\sm^2)_\mu\right)^4
+\left(\frac{1}{\Delta^2}(\sm^2)_\mu\frac{1}{\Delta^2}(\sm^2)_\nu\right)^2
+8\left(\frac{1}{\Delta^3}(\sm^2)_\mu\frac{1}{\Delta}(\sm^2)_\mu\right)^2
\nonumber \\ &&
+2\left(\frac{1}{\Delta^3}(\sm^2)_{\mu\mu}\right)^2
-8\frac{1}{\Delta^3}(\sm^2)_\mu\frac{1}{\Delta}(\sm^2)_\mu
\frac{1}{\Delta^3}(\sm^2)_{\nu\nu}
\nonumber \\ &&
+4\frac{1}{\Delta^2}(\sm^2)_\mu\frac{1}{\Delta^2}(\sm^2)_\nu
\frac{1}{\Delta^2}\sF_{\mu\nu}
+\left(\frac{1}{\Delta^2}\sF_{\mu\nu}\right)^2
\Bigg\}
\Bigg]\,.
\label{eq:40}
\end{eqnarray}
$2^{d/2}$ is the dimension of Dirac space and $\tr$ stands for trace
on flavor space only. Two further notational conventions have been
used. First, if $I$ is an ordered set of Lorentz indices $X_{\mu I}$
denotes\footnote{Note that in \cite{Caro:1993fs} this is denoted by
$X_{I\mu}$.}  $\sDa_\mu X_I$, thus, for instance,
\begin{equation}
\sm_\mu= \sDa_\mu\sm\,,\quad \sm_{\mu\nu}= \sDa_\mu(\sDa_\nu\sm)\,.
\end{equation}
Second, the bosonic determinant and all manipulations used are
invariant under the transformation defined by
\begin{equation}
\sm \mapsto \sm^\dagger=\sm\,,\quad 
\sD_\mu\mapsto \sD_\mu^\dagger=-\sD_\mu\,,\quad
(XY)^\dagger =Y^\dagger X^\dagger\,,
\end{equation}
(this is just the Hermitian conjugation when the fields have the
standard hermiticity). So, for instance,
\begin{equation}
\sm_\mu^\dagger=\sm_\mu\,,\quad
\sm_{\mu\nu}^\dagger=\sm_{\mu\nu}\,,\quad
\sF_{\mu\nu}^\dagger=-\sF_{\mu\nu}\,.
\end{equation}
Therefore, all terms in the previous formulas
either are self-conjugated or come in conjugated pairs. In writing the
formula for $W^+_4[v,m]$ each term has been identified with its 
conjugate \cite{Caro:1993fs,Muller:1996cq}. For instance
$\Delta^{-2}(\sm^2)_\mu\Delta^{-1}\sm_\nu \Delta^{-2}\sm_{\mu\nu}$,
actually stands for
\begin{equation}
\frac{1}{2}\frac{1}{\Delta^2}(\sm^2)_\mu\frac{1}{\Delta}\sm_\nu
\frac{1}{\Delta^2}\sm_{\mu\nu}
+
\frac{1}{2}  \frac{1}{\Delta^2} \sm_\nu
\frac{1}{\Delta} (\sm^2)_\mu 
\frac{1}{\Delta^2} \sm_{\mu\nu}\,.
\end{equation}
(Recall that terms related by a cyclic permutation are identified due
to the trace.)

In deriving the formulas for $W^+$ the identity
\begin{equation}
\left(\frac{1}{\Delta}\right)_\mu=
-\frac{1}{\Delta}(\sm^2)_\mu\frac{1}{\Delta}
\end{equation}
has been employed. In general it will not be practical to further
expand $(\sm^2)_\mu$ as $\sm\sm_\mu+\sm_\mu\sm$, since $\sm^2$ is a
chirality preserving quantity and so it is better behaved than $\sm$.
From the method used it follows that $W^+$ can be written in such a
way that single factors $\sm$ appear only through derivatives,
i.e. $\sm_\mu$, $\sm_{\mu\nu}$, etc, and underivated $\sm$ are always
squared.

\subsection{Momentum integrals}

Using the Convention 3, all momentum integrals in 
eqs.~(\ref{eq:43},\ref{eq:44},\ref{eq:40}) can explicitly be
computed. The ultraviolet divergences can be dealt with by using
dimensional regularization. The basic integral is of the form
\begin{equation}
I_{n,k,d}(z_1,\dots,z_n;\epsilon) =
\frac{(4\pi)^{\hat{d}/2}\Gamma(\hat{d}/2)}{(4\pi)^{d/2}\Gamma(d/2)}
\int\frac{d^{\hat{d}}p}{(2\pi)^{\hat{d}}}
\left(p^2\right)^r\prod_{i=1}^n\frac{1}{\Delta_i},
\label{eq:42}
\end{equation}
where
\begin{equation}
\hat{d}= d-2\epsilon\,, \quad k=\frac{d}{2}+r-1\,,\quad
\Delta_i= p^2+z_i^2\,.
\end{equation}
Here, $z_1,\dots,z_n$ are independent non vanishing variables (to be
replaced later by $\sm$). The integral can be done straightforwardly
and gives
\begin{equation}
I_{n,k,d} =
\frac{\Gamma(k-\epsilon+1)\Gamma(-k+\epsilon)}{(4\pi)^{d/2}\Gamma(d/2)}
\sum_{i=1}^n (z_i^2)^{k-\epsilon} R_{n,i} \,,
\end{equation}
with
\begin{equation}
R_{n,i}(z_1,\dots,z_n)= \prod_{j\not=i}\frac{1}{z_j^2-z_i^2}\,.
\end{equation}
Finally, expanding in powers of $\epsilon$ yields
\begin{equation}
I_{n,k,d}=
\frac{1}{(4\pi)^{d/2}\Gamma(d/2)} 
\left(
H_{n,k}+\left(\frac{1}{\epsilon}-\log(\mu^2)\right)Q_{n,k} + O(\epsilon)
\right) \,.
\label{eq:41}
\end{equation}
where the following functions have been introduced
\begin{eqnarray}
H_{n,k}(z_1,\dots,z_n;\mu) 
&=& (-1)^{k+1}\sum_{i=1}^n(z_i^2)^k\log(z_i^2/\mu^2)R_{n,i}\,,
\nonumber \\
Q_{n,k}(z_1,\dots,z_n)
 &=& (-1)^k\sum_{i=1}^n(z_i^2)^kR_{n,i}\,.
\label{eq:n49}
\end{eqnarray}

Several comments are in order here. 

(i) As indicated by the notation the functions $H_{n,k}$ and
$Q_{n,k}$ do not directly depend on the space-time dimension.

(ii) Following the standard practice, the parameter $\mu^2$ has been
introduced for convenience in the dimensional counting. The integral
$I_{n,k,d}$ does not depend on $\mu^2$ and so its dependence cancels in the
r.h.s. of \eq{41}. Under a scale transformation, $R_{n,i}$ and
$Q_{n,k}$ transform homogeneously, but
\begin{equation}
z_i^2\mapsto \lambda z_i^2\,,\quad H_{n,k} \mapsto
\lambda^{k+1-n}\left( H_{n,k} - \log(\lambda) Q_{n,k} \right) \,.
\end{equation}
Therefore, $Q_{n,k}$ represents the contribution of the integral to
the scale anomaly.

(iii) Whenever the integral $I_{n,k,d}$ is infrared and ultraviolet
convergent, the function $Q_{n,k}$ must vanish,
\begin{equation}
Q_{n,k}= 0 \quad \text{when}\quad 1\le k+1 <n \,.
\end{equation}
This is because as $\epsilon$ goes to zero $I_{n,k,d}$ has to remain
finite. In addition, $\log(\mu^2)$ no longer appears and there is no
scale anomaly. When the integral is infrared finite ($k\ge 0$, which
is the case of interest) but not necessarily ultraviolet finite, the
functions $Q_{n,k}$ are just homogeneous polynomials in the squared
masses, $z_i^2$, since taking a sufficient number of derivatives of
$I_{n,k,d}$ with respect to the any of the $z_i^2$, turns the integral
into a ultraviolet convergent one.

(iv) The ultraviolet finite prefactor
$\frac{(4\pi)^{\hat{d}/2}\Gamma(\hat{d}/2)}{(4\pi)^{d/2}\Gamma(d/2)}$
has been introduced in the definition of $I_{n,k,d}$ for convenience,
in order to remove renormalization dependent constant terms (the usual
$\log(4\pi)-\gamma$ of the MS scheme) in \eq{41}. Those constant
terms would come with $Q_{n,k}$ and would amount to further polynomial
contributions to the effective action.

(v) From the definition of $I_{n,k,d}$, it follows that $H_{n,k}$ and
$Q_{n,k}$ are completely symmetric functions of the $m^2_i$, and
furthermore, they remain finite as two or more of the squared masses
become equal. ($Q_{n,k}$ is a polynomial so the finiteness is obvious
in this case.)

(vi) The identity
\begin{equation}
\frac{1}{\Delta_i}\frac{1}{\Delta_j}
=-\frac{1}{z_i^2-z_j^2}\left(\frac{1}{\Delta_i}-\frac{1}{\Delta_j}\right)
\end{equation}
gives rise to the following recurrence relation
\begin{equation}
I_{n,k,d}(z_1,z_2,\dots;\epsilon)=
\frac{
I_{n-1,k,d}(z_2,z_3,\dots,z_n;\epsilon)
-I_{n-1,k,d}(z_1,z_3,\dots,z_n;\epsilon)
}{z_1^2-z_2^2} \,.
\end{equation}
As a consequence, analogous recurrence relations apply to $H_{n,k}$
and $Q_{n,k}$ as well, and they can be used to compute these functions
starting from
\begin{equation}
H_{1,k}= (-1)^{k+1}(z_1^2)^k\log(z_1^2/\mu^2) \,,\quad
Q_{1,k}= (-1)^k(z_1^2)^k \,.
\end{equation}

More generally, it will be necessary to consider also the case of
$\Delta_i$ raised to different powers in the momentum integral,
\eq{42}, namely
\begin{equation}
I_{n,k,d}^{r_1,\dots,r_n}(z_1,\dots,z_n;\epsilon)=
\frac{(4\pi)^{\hat{d}/2}\Gamma(\hat{d}/2)}{(4\pi)^{d/2}\Gamma(d/2)}
\int\frac{d^{\hat{d}}p}{(2\pi)^{\hat{d}}}
\left(p^2\right)^r\prod_{i=1}^n\frac{1}{\Delta_i^{r_i}}\,.
\end{equation}
The $r_i$ are assumed to be positive integers. Obviously, this
integral is given by the previous formulas by taking the first $r_1$
arguments to be all of them equal $z_1$, then the next $r_2$ arguments
to be $z_2$, and so on. That is,
\begin{equation}
I_{n,k,d}^{r_1,\dots,r_n}(z_1,\dots,z_n;\epsilon)=
I_{r_1+\cdots+r_n,k,d}(z_1,\dots,z_1,\dots,z_n,\dots,z_n;\epsilon)\,,
\end{equation}
where in the r.h.s. each $z_i$ appears $r_i$ times. The corresponding
analogous definitions apply to $H_{n,k}^{r_1,\dots,r_n}$ and
$Q_{n,k}^{r_1,\dots,r_n}$.

Alternatively, the same result is obtained taking derivatives with
respect to the $z_i^2$. For instance,
\begin{equation}
H_{n,k}^{r_1,\dots,r_n}(z_1,\dots,z_n;\mu)=
\prod_{i=1}^n\left[
\frac{(-1)^{r_i-1}}{(r_i-1)!}
\frac{\partial^{r_i-1}}{\partial(z_i^2)^{r_i-1}}
\right]
H_{n,k}(z_1,\dots,z_n;\mu)\,.
\end{equation}
Analogous formulas hold for $I_{n,k,d}^{r_1,\dots,r_n}$ and
$Q_{n,k}^{r_1,\dots,r_n}$.

It is interesting to note that $H_{n,k}$ has the form of a rational
function times a logarithm of $z_i^2/\mu^2$ (see the remark (iv)
above). This needs not be true for $H_{n,k}^{r_1,\dots,r_n}$; taking
the coincidence limit $z_i=z_j$ or derivating with respect to $z_i^2$
yields further terms which are purely rational functions of the
masses, without logarithms.

\subsection{Normal parity effective action through fourth order}

The momentum integrals in eqs.~(\ref{eq:43},\ref{eq:44},\ref{eq:40})
can straightforwardly be worked out using the previous formulas for
$I_{n,k,d}^{r_1,\dots,r_n}$. In the spirit of dimensional
regularization, the variable $d$ in
eqs.~(\ref{eq:43},\ref{eq:44},\ref{eq:40}) (excepting that in $d^dx$)
should be replaced by $\hat{d}=d-2\epsilon$ and a Laurent expansion in
$\epsilon$ is to be performed. After that there will be two types of
terms. First, contributions coming from
$(1/\epsilon-\log(\mu^2))Q_{n,k}^{r_1,\dots,r_n}$ in the momentum
integral. These pick up ultraviolet divergent terms plus finite terms
from the $\epsilon$ in $\hat{d}$. And second, finite terms from
$H_{n,k}^{r_1,\dots,r_n}$. The first kind of contributions are just
polynomials and so renormalization dependent.  Because all our
formulas are already manifestly gauge invariant before momentum
integration, those polynomials are gauge invariant and removable by
counter terms. (Within other schemes, the divergent polynomial terms
would still be gauge invariant but not necessarily the finite parts.)
Therefore modulo polynomials, the effective action can be obtained by
using $d$ instead of $\hat{d}$ and keeping only the finite part
$H_{n,k}^{r_1,\dots,r_n}$ in the momentum integrals.

The calculation of $W^+_0[v,m]$ in \eq{43} can be done by first taking
the derivative with respect to $\sm^2$. This removes the logarithm and
allows the application of the momentum integrals derived in the
previous subsection. $W^+_0[v,m]$ is then obtained by integrating back
with respect to $\sm^2$. This introduces an ambiguity which is just a
polynomial.
\begin{equation}
W^+_{0,d}[v,m]= 
-\frac{(-1)^{d/2}2^{d/2}}{2(4\pi)^{d/2}\Gamma(d/2+1)}
\int d^dx\,\tr\left[
\sm^d\log(\sm^2/\mu^2)\right] \,.
\label{eq:60a}
\end{equation}

Using our Convention 3 and the integrals of the previous section, the
expressions for $W^+_2[v,m]$ and $W^+_4[v,m]$ are a direct
transcription of eqs.~(\ref{eq:44},\ref{eq:40}). In the following
formulas the arguments $z_1,\dots,z_n$ of $H_{n,k}^{r_1,\dots,r_n}$
have to be substituted by $\sm_1,\dots,\sm_n$.
\begin{equation}
W^+_{2,d}[v,m]=
-\frac{1}{2} \frac{2^{d/2}}{(4\pi)^{d/2}\Gamma(d/2)}
\int d^dx \,\tr\left[
-\frac{1}{2} H_{2,d/2-1}^{1,1} \sm_\mu^2
+\frac{1}{d} H_{2,d/2}^{2,2} (\sm^2)_\mu^2
\right] \,.
\label{eq:60b}
\end{equation}
\begin{eqnarray}
W^+_{4,d}[v,m] &=&
-\frac{1}{2} \frac{2^{d/2}}{(4\pi)^{d/2}\Gamma(d/2)}
\int d^dx\,\tr\Bigg[
\nonumber \\ &&
-\frac{1}{2} H_{4,d/2-1}^{1,1,1,1} \sm_\mu^2 \sm_\nu^2
+\frac{1}{4} H_{4,d/2-1}^{1,1,1,1} (\sm_\mu \sm_\nu)^2
\nonumber \\ &&
+ H_{3,d/2-1}^{1,1,1} \sm_\mu \sm_\nu \sF_{\mu\nu}
+\frac{1}{4} H_{2,d/2-1}^{1,1} \sF_{\mu\nu}^2
\nonumber \\ &&
+\frac{2}{d}H_{4,d/2}^{2,1,2,1} ((\sm^2)_\mu \sm_\nu)^2
+\frac{2}{d}H_{4,d/2}^{2,2,1,1} ((\sm^2)_\mu \sm_\nu)^2
\nonumber \\ &&
+\frac{2}{d}H_{4,d/2}^{1,2,1,2} (\sm^2)^2_\mu \sm_\nu^2
+\frac{4}{d}H_{4,d/2}^{2,2,1,1} (\sm^2)^2_\mu \sm_\nu^2
\nonumber \\ &&
-\frac{4}{d}H_{3,d/2}^{2,1,2} (\sm^2)_\mu \sm_\nu\sm_{\mu\nu}
-\frac{4}{d}H_{3,d/2}^{2,2,1} (\sm^2)_\mu \sm_\nu\sm_{\mu\nu}
\nonumber \\ &&
+\frac{1}{d}H_{2,d/2}^{2,2} \sm_{\mu\nu}^2
\nonumber \\ &&
+\frac{4}{d(d+2)}H_{4,d/2+1}^{2,2,2,2} (\sm^2)_\mu^2 (\sm^2)_\nu^2 
-\frac{2}{d(d+2)}H_{4,d/2+1}^{2,2,2,2} ((\sm^2)_\mu (\sm^2)_\nu)^2 
\nonumber \\ &&
-\frac{16}{d(d+2)}H_{4,d/2+1}^{3,1,3,1} (\sm^2)_\mu^2 (\sm^2)_\nu^2 
-\frac{4}{d(d+2)}H_{2,d/2+1}^{3,3} (\sm^2)_{\mu\mu}^2
\nonumber \\ &&
+\frac{16}{d(d+2)}H_{3,d/2+1}^{3,1,3} (\sm^2)_\mu^2 (\sm^2)_{\nu\nu}
\nonumber \\ &&
-\frac{8}{d(d+2)}H_{3,d/2+1}^{2,2,2} (\sm^2)_\mu (\sm^2)_\nu \sF_{\mu\nu}
-\frac{2}{d(d+2)}H_{2,d/2+1}^{2,2}\sF_{\mu\nu}^2
\Bigg] \,.
\label{eq:60}
\end{eqnarray}

\section{Discussion}
\label{sec:3}

\subsection{Expression in terms of matrix elements}

In order to  analyze the previous formulas, let us consider in
more detail the second order term in two space-time dimensions,
$W^+_{2,2}[v,m]$. A straightforward calculation yields
\begin{eqnarray}
W^+_{2,2}[v,m] &=& \frac{1}{8\pi}\int d^2x\,\tr\,\Bigg[
H(\sm_1,\sm_2) \sm_\mu^2
-\frac{(\sm_1^2+\sm_2^2)H(\sm_1,\sm_2)-2}
{(\sm_1^2-\sm_2^2)^2} (\sm^2)_\mu^2
\Bigg] \,,
\label{eq:17b}
\end{eqnarray}
where
\begin{equation}
H(x,y)= \frac{\log(x^2/y^2)}{x^2-y^2}\,.
\end{equation}
The symmetry under the exchange of the ordering labels 1,2 is a direct
consequence of the cyclic property of the trace.  

As noted, this kind of formulas can be evaluated using basis of
eigenvectors of $\sm^2$ at each point $x$. This gives (Convention 1 is
still at work)
\begin{eqnarray}
W^+_{2,2}[v,m] &=& \frac{1}{8\pi}\int d^2x\,\sum_{j,k}\,
\Bigg[
H(m_j,m_k) (\Da_\mu\mrl)_{jk}(\Da_\mu\mlr)_{kj}
\nonumber \\
&& - \frac{ (m_j^2+m_k^2)H(m_j,m_k)-2 }{(m_j^2-m_k^2)^2}
(\Da_\mu m^2_R)_{jk} (\Da_\mu m^2_R)_{kj}
\Bigg] \,.
\label{eq:19a}
\end{eqnarray}
In this formula $(\Da_\mu\mrl)_{jk}= \langle j,R|\Da_\mu\mrl|
k,L\rangle$, etc. The quantities $m^2_{R,L}$ were defined in \eq{n6}.

Particular attention requires the case of $\sm_1^2$ and $\sm^2_2$ with
the same eigenvalue. Before momentum integration it is clear that this
case is a perfectly regular one. It follows that the correct result
after momentum integration can be obtained by taking the finite formal
limit $\sm^2_2\to\sm^2_1$, and then replacing $\sm^2$ by its
eigenvalue. That is, in \eq{19a} the case $j=k$ (or more generally,
$m_j^2 = m_k^2$) is resolved by taking the limit $m_k^2\to m_j^2$
which is well defined and finite, namely
\begin{eqnarray}
&& \frac{1}{8\pi}\int d^2x\,\sum_j\, \left( 
\frac{1}{m_j^2} (\Da_\mu\mrl)_{jj}(\Da_\mu\mlr)_{jj}
-\frac{1}{6m_j^4} (\Da_\mu m^2_R)^2_{jj}
\right) 
\nonumber \\
&=&
\frac{1}{8\pi}\int d^2x\,\sum_j\, \left( 
\frac{2}{3m_j^2} (\Da_\mu\mrl)_{jj}(\Da_\mu\mlr)_{jj}
-\frac{1}{3m_j^2} (\Da_\mu\mrl)^2_{jj}
\right) \,.
\label{eq:18a}
\end{eqnarray}
This is the full result in the Abelian case.

As an application, consider $\mlr(x)= M(x)U(x)$, $\mrl(x)=
M(x)U^{-1}(x)$, where $M(x)$ is a c-number but not necessarily
constant. Since $\sm^2=M^2$ is c-number, the calculation can be done
directly from \eq{17b} by taking $\sm^2=M^2$ everywhere. This gives
\begin{equation}
W^+_{2,2} = \frac{1}{4\pi}\int d^2x\,\tr\,\left[
\frac{1}{6}\left(\frac{\partial_\mu M}{M}\right)^2
-\frac{1}{2} (U^{-1}\Da_\mu U)^2\right]\,.
\end{equation}

\subsection{Commutator expansions}
\label{subsec:4.a}

The derivative expansion treats $m$ non perturbatively. In other
approaches the result appears instead in the form of a $1/m$-like
expansion such that each term is an homogeneous function of $m$
\cite{Ball:1986qr,Ball:1989xg}. The order of each term in such an
expansion is counted by the number of commutators (including those
implied by the covariant derivative). Therefore, when operators $\sm$
in the middle of an expression are moved to left, producing
commutators, the new terms so generated count as higher order than the
original one and hence they are suppressed in this counting. Although
the present work is devoted to derivative expansions, in this
subsection we will consider commutator expansion of our formulas.

Using our Convention 3 it is remarkably simple to systematically
reexpand $W^+$ in terms of commutators. This can be illustrated with
$W^+_{2,2}$ in form given in \eq{13}  (this expression will be derived
below). From \eq{n25} it follows that the quantity 
\begin{equation}
\scom_1=\sm_1-\sm_2
\end{equation}
is equivalent to a commutator $[\sm,\ ]$ on the first $\sm_\mu$. For
instance,
\begin{equation}
\scom_1^2 \sm_\mu^2= (\sm_1-\sm_2)^2 \sm_\mu^2 =
(\sm_1-\sm_2)[\sm,\sm_\mu]\sm_\mu = [\sm,[\sm,\sm_\mu]]\sm_\mu \,.
\end{equation}
Therefore, a commutator expansion is obtained by removing the variable
$\sm_2$ in favor of $\scom_1$, i.e., $\sm_2=\sm_1-\scom_1$, and
carrying out an expansion in powers of $\scom_1$:
\begin{equation}
-\frac{\sm_1\sm_2 H(\sm_1,\sm_2) -1}{(\sm_1-\sm_2)^2}
=\frac{1}{6}\frac{1}{\sm_1^2} 
+\frac{1}{6}\frac{\scom_1}{\sm_1^3}
+\frac{2}{15}\frac{\scom_1^2}{\sm_1^4}
+\cdots\,.
\label{eq:16}
\end{equation}
This immediately translates into
\begin{eqnarray}
W^+_{2,2}[v,m] &=& \frac{1}{4\pi}\int d^2x\,\tr\,\left[
\frac{1}{6\sm^2} \sm_\mu^2 
+\frac{1}{6\sm^3} [\sm,\sm_\mu]\sm_\mu 
+\frac{2}{15\sm^3} [\sm,[\sm,\sm_\mu]]\sm_\mu +\cdots
\right] \,.
\end{eqnarray}

For $W^+_4$ (or higher orders) the identity $\scom_n= \sm_n-\sm_{n+1}$
can be used, where $\scom_n$ denotes the commutator $[\sm,\ ]$ on the
$n$-th fixed element.

Due to the cyclic property (and in particular integration by parts
when $\sD_\mu$ is involved) neither the derivative expansion nor the
commutator expansion take a unique form. However, there is an
important difference between both expansions regarding their
uniqueness, namely, the derivative expansion is related to (gauge
covariant) dilatations of the external fields $v$ and $m$; each order
is tied to a given power of the dilatation parameter. This guarantees
that each order in the derivative expansion is a well-defined
functional of $v$ and $m$. (That is, each given order can be written
in different forms but has a single numerical value for each given
configuration of the fields.) On the other hand, the commutator
expansion is not tied to any expansion parameter and so it is possible
to have two different commutators expansions of a single functional
which differ numerically at every order. For instance, in the identity
\begin{equation}
\tr(2\sm[\sm,\sm_\mu]\sm_\mu) = \tr([\sm,[\sm,\sm_\mu]]\sm_\mu) 
\end{equation}
the same functional is represented by two terms of different order
(i.e., with a different number of commutators), even imposing a
standard form ($\sm_1$ and $\scom_1$ have been chosen as the
independent variables). More generally, a functional of the form
$\tr(f(\sm_1,\sm_2)\sm_\mu^2)$ depends only on the symmetric component
of the function $f(x_1,x_2)$, but the antisymmetric component may have
a non vanishing commutator expansion when $\sm_1$ and $\scom_1$ are
used as the independent variables. This suggest to first symmetrize
$f$ and then expand.\footnote{What seems to be true is that, for a
given non vanishing functional, the order of the leading term (i.e.,
the number of commutators in the term with the least number of
commutators) has an upper bound over the set of all possible
commutators expansions of the functional.} In any case, it is clear
the usefulness for our notation to analyze this kind of problems and
even more so for more complicated cases, such as contributions to
$W^+_4$ or higher orders.

The ambiguity in the commutator expansion allows to reorder
$W_{2,2}^+$ so that the symmetry under the exchange of ordering labels
1,2 in the right-hand side of \eq{16} is restored.  A convenient
choice, containing even orders only, is
\begin{equation}
-\frac{\sm_1\sm_2 H(\sm_1,\sm_2) -1}{(\sm_1-\sm_2)^2}
= \frac{1}{6}\frac{1}{\sm_1\sm_2}
+\frac{1}{30}\frac{\scom_1}{\sm_1^2}\frac{\scom_2}{\sm_2^2}
+\frac{1}{140}\frac{\scom_1^2}{\sm_1^3}\frac{\scom_2^2}{\sm_2^3}
+\cdots\,,
\label{eq:c71}
\end{equation}
where $\scom_2=\sm_2-\sm_3= -\scom_1$ represents $[\sm,\ ]$ on the
second $\sm_\mu$. This expansion can be obtained from that in \eq{16}
by recursively subtracting the symmetric version of the leading term
and then symmetrizing the remainder. Eq. (\ref{eq:c71}) translates
into
\begin{eqnarray}
W^+_{2,2}[v,m] &=&  \frac{1}{4\pi}\int
d^2x\,\tr\,\left[ \frac{1}{6}\left(\frac{1}{\sm}\sm_\mu\right)^2
+\frac{1}{30}\left(\frac{1}{\sm^2}[\sm,\sm_\mu]\right)^2
+\frac{1}{140}\left(\frac{1}{\sm^3}[\sm,[\sm,\sm_\mu]]\right)^2
+\cdots \right] \,.
\label{eq:n50}
\end{eqnarray}

The commutator expansion just derived is well suited for the
vector-like case. In many cases, however, an expansion involving
commutators of $\sm^2$ will be more useful. For instance, in the
Abelian case $\sm^2$ is always a (not necessarily constant) c-number
whereas $\sm$ is not (i.e. $[\sm,\sX]$ needs not vanish even when
$\sX$ is a multiplicative operator). The method to obtain such an
expansion is the same as above and can be applied to $W^+_{2,2}$ in
\eq{17b}. Namely, the quantity 
\begin{equation}
\sC_1= \sm_1^2-\sm_2^2
\end{equation}
represents $[\sm^2,\ ]$ applied to the first fixed factor ($\sm_\mu$
or $(\sm^2)_\mu$) thus $\sm_2^2$ can eliminated in favor of $\sC_1$ in
\eq{17b}. This gives
\begin{eqnarray}
H(\sm_1,\sm_2) &=&
\frac{1}{\sm_1^2}
+\frac{\sC_1}{2\sm_1^4}
+\frac{\sC_1^2}{3\sm_1^6}
+\frac{\sC_1^3}{4\sm_1^8}
+\frac{\sC_1^4}{5\sm_1^{10}}
+\cdots \,,
\nonumber \\
\frac{(\sm_1^2+\sm_2^2) H(\sm_1,\sm_2) - 2}{(\sm_1^2-\sm_2^2)^2} &=&
\frac{1}{6\sm_1^4}
+\frac{\sC_1}{6\sm_1^6}
+\frac{3\sC_1^2}{20\sm_1^8}
+\frac{2\sC_1^3}{15\sm_1^{10}}
+\frac{5\sC_1^4}{42\sm_1^{12}}
+\cdots \,.
\end{eqnarray}
Further, applying a (in this case partial) symmetrization in the
labels 1,2 to remove terms with an odd number of commutators, yields
\begin{eqnarray}
H(\sm_1,\sm_2) &=& \frac{1}{\sm_1^2}\left(
1
+\frac{1}{6}\frac{\sC_1\sC_2}{\sm_1^2\sm_2^2}
+\frac{1}{30}\frac{\sC_1^2\sC_2^2}{\sm_1^4\sm_2^4}
+\cdots  \right) \,,
\nonumber \\
\frac{(\sm_1^2+\sm_2^2) H(\sm_1,\sm_2) - 2}{(\sm_1^2-\sm_2^2)^2} &=&
\frac{1}{6}\frac{1}{\sm_1^2\sm_2^2}
+\frac{1}{60}\frac{\sC_1\sC_2}{\sm_1^4\sm_2^4}
+\frac{1}{420}\frac{\sC_1^2\sC_2^2}{\sm_1^6\sm_2^6}
+\cdots \,.
\end{eqnarray}
(Where $\sC_2= \sm_2^2-\sm_3^2=-\sC_1$.) This translates into
\begin{eqnarray}
W^+_{2,2}[v,m] &=& \frac{1}{8\pi}\int d^2x\,\tr\,\Bigg[
\frac{1}{\sm^2}\left(
\sm_\mu^2
+\frac{1}{6}\left(\frac{1}{\sm^2}[\sm^2,\sm_\mu]\right)^2
+\frac{1}{30}\left(\frac{1}{\sm^4}[\sm^2,[\sm^2,\sm_\mu]]\right)^2
+\cdots \right)
\nonumber \\ 
&&
-\frac{1}{6}\left(\frac{1}{\sm^2}(\sm^2)_\mu\right)^2
-\frac{1}{60}\left(\frac{1}{\sm^4}[\sm^2,(\sm^2)_\mu]\right)^2
-\frac{1}{420}\left(\frac{1}{\sm^6}[\sm^2,[\sm^2,(\sm^2)_\mu]]\right)^2
+\cdots 
 \Bigg]  \,.
\end{eqnarray}

In fully expanded notation the leading term is (compare with \eq{18a})
\begin{eqnarray}
W^+_{2,2,\text{leading}}[v,m] &=& \frac{1}{16\pi}\int d^2x\,\tr\,\Bigg[
\frac{1}{m^2_R}\Da_\mu\mrl \Da_\mu\mlr
+\frac{1}{m^2_L}\Da_\mu\mlr \Da_\mu\mrl
\nonumber \\
&& 
-\frac{1}{6}\left(\frac{1}{m^2_R}\Da_\mu m^2_R \right)^2
-\frac{1}{6}\left(\frac{1}{m^2_L}\Da_\mu m^2_L \right)^2
\Bigg]\,.
\end{eqnarray}
In the example considered above, $\sm^2=M^2$ a (non necessarily
constant) c-number, this is the full contribution.

\subsection{Alternative expressions}

The expression in \eq{17b} can be somewhat simplified by using the
identity (valid inside the trace)
\begin{equation}
(\sm^2)_\mu^2= \{\sm,\sm_\mu\}^2=(\sm_1+\sm_2)(\sm_2+\sm_3)\sm_\mu^2
=(\sm_1+\sm_2)^2\sm_\mu^2
\end{equation}
(recall that $\sm_3=\sm_1$ due to the cyclic property). This yields
the formula
\begin{equation}
W^+_{2,2}[v,m] = -\frac{1}{4\pi}\int d^2x\,\tr\left[
\frac{\sm_1\sm_2H(\sm_1,\sm_2) -1}{(\sm_1-\sm_2)^2} \sm_\mu^2 \right]
\,.
\label{eq:13}
\end{equation}
In most instances, in order to use this kind of formulas it should be
clear which are the chiral labels $L,R$ in each of the factors. The
simplest way to determine this is by forcing that the ordering labels
1,2 appear only in $\sm^2$, since this object is either $RR$ or $LL$
and so does not flip the chirality label. This can always be achieved
by splitting the expression into components which are even or odd
under $\sm_1\to -\sm_1$ and $\sm_2\to -\sm_2$, that is, in an
expression of the form $\tr[f(\sm_1,\sm_2)\sm_\mu^2]$
\begin{equation}
f(\sm_1,\sm_2)=
A(\sm_1^2,\sm_2^2)+\sm_1\sm_2B(\sm_1^2,\sm_2^2)\,.
\label{eq:24}
\end{equation}
(Note that any expression must be even under $\sm\to -\sm$ since there
should be the same number of $R$ and $L$ labels. Therefore no terms of
the form $\sm_1D(\sm_1^2,\sm_2^2)+\sm_2F(\sm_1^2,\sm_2^2)$ can
appear.)  Applying this procedure to \eq{13} yields
\begin{eqnarray}
W^+_{2,2}[v,m] &=& -\frac{1}{4\pi}\int d^2x\,\tr\,\Bigg[
\frac{(\sm_1^2+\sm_2^2)H(\sm_1,\sm_2)-2} {(\sm_1^2-\sm_2^2)^2}
(\sm\sm_\mu)^2 \nonumber \\ &&
+\frac{2\sm_1^2\sm_2^2H(\sm_1,\sm_2)-(\sm_1^2+\sm_2^2)}
{(\sm_1^2-\sm_2^2)^2} \sm_\mu^2 \Bigg] \,.
\label{eq:17}
\end{eqnarray}
Now, from our conventions it unambiguously follows that in the first
term the chiral labels of $\sm_1^2$ and $\sm_2^2$ as well as of both
factors $(\sm\sm_\mu)$ are $RR$. In the second term, $\sm_1^2$ is
$RR$, $\sm_2^2$ is $LL$, the first factor $\sm_\mu$ is $RL$ and
the second one is $LR$. (Recall that under the Convention 1, each term
is identified with its pseudo-parity conjugate.) This allows to take
matrix elements as in \eq{19a} in a straightforward manner. (The
original expression \eq{17b} was already in this form.)  The procedure
expressed by \eq{24} shows that any expression, such as that in
\eq{13}, can be brought to a standard form, \eq{17}, which is free of
ambiguities, i.e., the expression can be expanded undoing the
Conventions 1 and 2 in an unambiguous way. This procedure immediately
extends to more than two variables, as required in $W^+_4$ or higher
order terms.

An essential point in this discussion is that, in an expression
$\tr(f(\sm_1,\sm_2)\sm_\mu^2)$, the function $f(x_1,x_2)$ must be
regular in the coincidence limits. To further analyze this point,
consider the contributions to $W^+_{2,2}$ in \eq{13} coming solely
from $\sv_\mu$, i.e. setting $\partial_\mu$ to zero (recall that
$\sD_\mu=\partial_\mu+\sv_\mu$). Using the identity 
\begin{equation}
[\sv_\mu,\sm]^2= -(\sm_1-\sm_2)^2\sv_\mu^2 \,,
\end{equation}
which holds inside the trace, this gives
\begin{equation}
W^+_{2,2}[v,m] = \frac{1}{4\pi}\int d^2x\,\tr\left[
\left(\sm_1\sm_2H(\sm_1,\sm_2) -1\right) \sv_\mu^2 \right]
+O(\partial)\,.
\label{eq:13a}
\end{equation}
Because the action depends exclusively on the combination
$\sD=\partial+\sv$ and no algebraic assumption has been made on $\sv$
(it can be an arbitrary matrix), it is clear that all the effective
action can be reconstructed from the case $\partial_\mu=0$ by means of
the replacement $\sv\to \sD$ everywhere (this a reciprocal of the
usual gauging procedure). So in a formal sense
\begin{equation}
W^+_{2,2}[v,m] = \frac{1}{4\pi}\int d^2x\,\tr\left[
\left(\sm_1\sm_2H(\sm_1,\sm_2) -1\right) \sD_\mu^2 \right] \,.
\label{eq:13b}
\end{equation}
This is meaningful provided that $\sD_\mu$ (or $\sv_\mu$ prior to the
``gauging'') appears only in commutators. Under our Convention 3, this
can be easily enforced through the formal identity
\begin{equation}
\sD_\mu^2 = - \frac{\sm_\mu^2}{(\sm_1-\sm_2)^2} \,.
\label{eq:1}
\end{equation}
The minus sign comes because $\sm_\mu= \sD_\mu\sm- \sm\sD_\mu=
(\sm_2-\sm_1)\sD_\mu$ for the first $\sD_\mu$ in \eq{13b}, whereas
$\sm_\mu=(\sm_3-\sm_2)\sD_\mu=(\sm_1-\sm_2)\sD_\mu$ for the second
one. This immediately recovers the correct expression \eq{13}.

This does not mean however that the formal procedure is always
justified. In fact, we could have considered, instead of \eq{1}, a
different formal expression, namely
\begin{equation}
\sD_\mu^2 = - \frac{(\sm^2)_\mu^2}{(\sm_1^2-\sm_2^2)^2} \,.
\label{eq:1a}
\end{equation}
This can be used in \eq{13b} and the result can be rewritten (in the
spirit of \eq{24}) as
\begin{equation}
W^+_{2,2}[v,m] = -\frac{1}{4\pi}\int d^2x\,\tr\,
\frac{1}{(\sm_1^2-\sm_2^2)^2}
\left[
H(\sm_1,\sm_2)(\sm(\sm^2)_\mu)^2 -(\sm^2)_\mu^2
\right] \,.
\label{eq:81}
\end{equation}
This expression is incorrect if taken naively.  Because it involves
derivatives of $\sm^2$ only, it would predict, for instance, a
vanishing value for $W^+_{2,2}$ when $m$ is on the chiral circle,
i.e. when $\sm^2$ is a constant c-number, or that $W^+_{2,2}$ does not
depend on the axial field in the Abelian case, both predictions being
wrong. The reason is that Eq.~(\ref{eq:1}) introduces a singularity at
$\sm_1=\sm_2$ which is canceled by the numerator
$\sm_1\sm_2H(\sm_1,\sm_2) -1$, but \eq{1a} introduces a new
singularity at $\sm_1=-\sm_2$ which is not canceled; each of the terms
in \eq{81} is not separately finite when $\sm_1^2=\sm_2^2$ and in
general they cannot cancel to each other due to their different chiral
labels. This renders the expression meaningless unless $(\sm^2)_\mu$
is expanded to cancel the spurious singularity at $\sm_1=-\sm_2$.

The fact that the formal expression \eq{13b} can be promoted to a
regular expression with $\sD_\mu$ in commutators (cf. \eq{13}) is a
manifestation of gauge invariance. It is perhaps interesting to note
that the gauge invariance of \eq{13b} can be checked even without
actually bringing $\sD_\mu$ into commutators. Namely, $\sD_\mu$
appears only in commutators in an expression, if and only if, the
expression remains invariant under the shift
$\sD_\mu\to\sD_\mu+a_\mu$, where $a_\mu$ is an arbitrary constant
c-number. That the formula \eq{13b} is gauge invariant can be seen by
applying this test: it is readily seen that in the terms with $a_\mu$,
$\sm_1$ and $\sm_2$ coincide (since $\sm$ and $a_\mu$ commute) and the
limit $\sm_1\to\sm_2$ gives zero. (That it is a double zero follows
from symmetry under exchange of labels 1,2.)

Finally, let us mention that \eq{13a}, and thus \eq{13b}, can be
obtained rather directly in several ways without using Chan's
method. The key point is that in the absence of $\partial_\mu$ all
quantities are multiplicative operators and the cyclic property holds.
The procedure implied by combining \eq{13b} and \eq{1}, can then be
extended to compute $W^+_2[v,m]$ in any number of dimensions. However
this method is not suitable to treat the case of four or more
derivatives. For instance, the formula similar to \eq{13b} to fourth
order is
\begin{eqnarray}
W^+_{4,2} &=& 
\frac{1}{4\pi}\int d^2x\,\tr\,\Big[
\left((\sm_1\sm_3-\sm_2\sm_4)H_{4,1}+\sm_1\sm_2\sm_3\sm_4H_{4,0}\right)
\sD_\mu\sD_\mu\sD_\nu\sD_\nu 
\nonumber \\
&& -\frac{1}{2}\left(
(\sm_1\sm_3+\sm_2\sm_4)H_{4,1}+\sm_1\sm_2\sm_3\sm_4H_{4,0}\right)
\sD_\mu\sD_\nu\sD_\mu\sD_\nu
\Big]\,,
\end{eqnarray}
where $H_{4,k}$ are the functions of $\sm_1,\sm_2,\sm_3,\sm_4$ defined
in \eq{n49}. Using their explicit form and the test noted above of
shifting $\sD_\mu$ by a constant c-number, it is possible to check
that the previous formula is chiral invariant (of course, this is just
a check of the calculation). However, there is no simple systematic
form for bringing it to an explicitly invariant form where all
covariant derivatives appear only in commutators. The trick introduced
in \eq{1} works in $W^+_2$ because $\sD_\mu$ appears there as a first
covariant derivative. Such replacement is no longer sufficient in
$W^+_4$ since explicit chiral invariance requires at least the
presence of second covariant derivatives as well as
$\sF_{\mu\nu}$. (If only the replacement in \eq{1} is used, one
obtains an expression involving $\sm_\mu$ with coefficient which are
functions of $\sm_1,\sm_2,\sm_3,\sm_4$. However, these functions are
not finite in the coincidence limit, i.e. when two or more of their
argument become equal, and such an expression is not truly
well-defined.)

\subsection{Bosonic formulas}

For completeness we give here the analogous formulas of
eqs. (\ref{eq:43},\ref{eq:44},\ref{eq:40}) and
eqs. (\ref{eq:60a},\ref{eq:60b},\ref{eq:60}) for the bosonic effective
action.
\begin{eqnarray}
\Tr\log(P^2+U) &=&
\int\frac{d^dx\,d^dp}{(2\pi)^d}\tr \Bigg[ 
\log\Delta
+\frac{p^2}{d}\left(\frac{1}{\Delta^2}U_\mu\right)^2
\nonumber \\ &&
-\frac{2p^4}{d(d+2)}\Bigg\{
-2\left(\frac{1}{\Delta^2}U_\mu\right)^4
+\left(\frac{1}{\Delta^2}U_\mu\frac{1}{\Delta^2}U_\nu\right)^2
+8\left(\frac{1}{\Delta^3}U_\mu\frac{1}{\Delta}U_\mu\right)^2
\nonumber \\ &&
+2\left(\frac{1}{\Delta^3}U_{\mu\mu}\right)^2
-8\frac{1}{\Delta^3}U_\mu\frac{1}{\Delta}U_\mu
\frac{1}{\Delta^3}U_{\nu\nu}
\nonumber \\ &&
+4\frac{1}{\Delta^2}U_\mu\frac{1}{\Delta^2}U_\nu
\frac{1}{\Delta^2}F_{\mu\nu}
+\left(\frac{1}{\Delta^2}F_{\mu\nu}\right)^2
\Bigg\}
+ \cdots
\Bigg]\,,
\end{eqnarray}
where $\Delta=p^2+U$, $U_\mu=[P_\mu,U]$, etc. From this formula, the
fermionic one is obtained by setting $U=\sm^2+U^{(1)}+U^{(2)}$.

After momentum integration, the renormalized bosonic effective action
becomes
\begin{eqnarray}
\Tr\log(P^2+U) &=&
\frac{1}{(4\pi)^{d/2}\Gamma(d/2)}
\int d^dx\,\tr\Bigg[
(-1)^{d/2}\frac{2}{d}U^{d/2}\log(U/\mu^2)
+\frac{1}{d} H_{2,d/2}^{2,2} U_\mu^2
\nonumber \\ &&
+\frac{4}{d(d+2)}H_{4,d/2+1}^{2,2,2,2} U_\mu^2 U_\nu^2 
-\frac{2}{d(d+2)}H_{4,d/2+1}^{2,2,2,2} (U_\mu U_\nu)^2 
\nonumber \\ &&
-\frac{16}{d(d+2)}H_{4,d/2+1}^{3,1,3,1} U_\mu^2 U_\nu^2 
-\frac{4}{d(d+2)}H_{2,d/2+1}^{3,3} U_{\mu\mu}^2
+\frac{16}{d(d+2)}H_{3,d/2+1}^{3,1,3} U_\mu^2 U_{\nu\nu}
\nonumber \\ &&
-\frac{8}{d(d+2)}H_{3,d/2+1}^{2,2,2} U_\mu U_\nu F_{\mu\nu}
-\frac{2}{d(d+2)}H_{2,d/2+1}^{2,2}F_{\mu\nu}^2
+ \cdots
\Bigg] \,.
\end{eqnarray}
In this case, the arguments $z_1^2,\dots,z_n^2$ in
$H_{n,k}^{r_1,\dots,r_n}$ have to be substituted by $U_1,\dots,U_n$.

\section*{Acknowledgments}
This work was supported in part by funds provided by the Spanish
DGICYT grant no. PB98-1367 and Junta de Andaluc\'{\i}a grant
no. FQM-225.

\end{document}